\documentclass[a4paper,12pt]{article}

\begin{document}

\author{C. Barrab\`es\thanks{E-mail : barrabes@celfi.phys.univ-tours.fr}
 \\     
\small Laboratoire de Math\'ematiques et Physique Th\'eorique,\\
\small  CNRS/UMR 6083, Universit\'e F. Rabelais, 37200 TOURS, 
France\\  
P.A. Hogan\thanks{E-mail : phogan@ollamh.ucd.ie}\\
\small Mathematical Physics Department,\\
\small  National University of Ireland Dublin, Belfield, Dublin 4, Ireland and 
\\W. Israel\thanks{E-mail : israel@uvic.ca}\\
\small Physics Department,\\
\small University of Victoria, Victoria, Canada}

\title{The Aichelburg--Sexl Boost of Domain-Walls and Cosmic Strings}
\date{}
\maketitle

\begin{abstract}

We consider the application of the Aichelburg-Sexl boost to
plane and line distributions of matter. Our analysis shows
that for a domain-wall the space--time after the boost is flat
except on a null hypersurface which is the history of a null
shell. For a cosmic string we study the influence of the boost
on the conical singularity and give the new value of the
conical deficit. 

\end{abstract}
\thispagestyle{empty}
\newpage

\section{Introduction}\indent

The properties of the spacetime which results from boosting a spherical mass
to the velocity of light were first studied by Aichelburg and Sexl \cite{AS}.
This was later extended to the Kerr geometry \cite{Bal}-\cite{Mag} 
and to a static axially symmetric source characterized by its 
multipole moments \cite{BH1}. 
In all those cases the space--time after the boost is flat except on a  
singular null hypersurface. The Riemann tensor vanishes everywhere
except on this surface, its non vanishing components have 
a $\delta$-type singularity whose value depends on the mass 
of the boosted particle \cite{Bal} or on the multipole moments of
the boosted source \cite{BH1}. Furthermore the singular null hypersurface 
which is produced by the boost represents the history of
a plane impulsive gravitational wave and admits a singular
line along one of its null generators.
In all these examples the Aichelburg-Sexl boost produces a change of the
algebraic type of the Weyl tensor.

The properties of singular null hypersufaces, i.e. spacetimes
with a Riemann tensor having a $\delta$-type singularity with
support on a null hypersurface, have been studied already 
\cite{BI}-\cite{BH2}. It has been shown that in general
a null shell and an impulsive gravitational wave co-exist
and that the singular null hypersurface represents the
history of both the shell and the wave.  
One may then ask the following question: Can one find sources
for which the Aichelburg-Sexl boost  
produces a null shell in addition to, or in place of, 
an impulsive gravitational wave?

In this work we consider two examples, one corresponding to a plane 
distribution of matter and the other to a
line source. Planar sources have received considerable attention
in the recent past in the context of brane-cosmology. 
We have chosen here to study the effect of the Aichelburg-Sexl boost
on planar domain-walls both in vacuum and in (Anti) de Sitter space,
and with arbitrary dimension.
Generalization to walls with different equations of state
would be straightforward from our analysis.
We also consider the case of a 
straight cosmic string. An interesting aspect of this  case is that the initial
space--time possesses a conical singularity and one may wonder how this
geometrical property is affected by the boost.

The Aichelburg-Sexl boost has been utilised in the study of the 
dynamics of compact objects at extreme energies such as for instance
the scattering of ultra-relativistic black holes \cite{DE}.
More recently it has received much attention in a classical model
for high-energy scattering in quantum gravity which offers the 
exciting prospect that black holes could be produced in
future accelerators \cite{Gid}.

%\newpage
     
\setcounter{equation}{0}
\section{Boosting Domain Walls}\indent

Domain walls are two--dimensional topological defects in ordinary space and 
are characterised by an equation of state for which the surface energy density 
is equal to minus the surface pressure. The space--time model in  
general relativity of the gravitational field of such a source 
has been given by Vilenkin \cite{Vil} for planar walls. It is 
conformally flat, locally flat (except on the hypersurface $z=0$, the 
history of the wall) and all slices $z={\rm constant}$ are isometric to 
de Sitter space.

We first generalise the Vilenkin solution to $(n+1)$--dimensional space--time $M$. 
The domain wall is thus an $(n-1)$--dimensional surface and its history is 
a time--like $n$--dimensional hypersurface in $M$. We will consider only 
planar domain walls and start with a line--element for $M$ having the form
\begin{equation}\label{2.1}
ds^2=g_{\alpha\beta}\,dx^{\alpha}dx^{\beta}={\rm e}^{2V(z)}\,
\gamma _{ab}(x^c)\,dx^a\,dx^b+dz^2\ .
\end{equation} 
Here Greek indices take values $0, 1, 2,\dots , n$ while Latin indices take values 
$0, 1, 2,\dots ,n-1$ and $x^n=z$. Also $V$ is a function of $z$ only and $\gamma _{ab}=\gamma _{ba}$ 
are functions of $x^c$ only. The components of the Riemann curvature tensor 
for the metric given via the line--element (\ref{2.1}) are
\begin{eqnarray}\label{2.2}
R_{abcd}&=&{\rm e}^{2V}\,R_{abcd}(\gamma )+{\rm e}^{4V}\,V'^2(\gamma _{ad}
\,\gamma _{bc}-\gamma _{ac}\,\gamma _{bd})\ ,\\
R_{abcn}&=&0\ ,\\
R_{ancn}&=&-{\rm e}^{2V}(V''+V'^2)\,\gamma _{ac}\ ,
\end{eqnarray}
where $R_{abcd}(\gamma )$ are the components of the Riemann curvature tensor 
calculated with the metric tensor $\gamma _{ab}$. The Ricci tensor components 
for the metric tensor $g_{\alpha\beta}$ given above are
\begin{eqnarray}\label{2.5}
R_{ab}&=&R_{ab}(\gamma )-{\rm e}^{2V}\,(V''+n\,V'^2)\,\gamma _{ab}\ ,\\
R_{an}&=&0\ ,\\
R_{nn}&=&-n\,(V''+V'^2)\ .
\end{eqnarray}
If we now require that the space--time $M$ be conformally flat then the vanishing 
of the Weyl tensor for the metric $g_{\alpha\beta}$ results in 
\begin{equation}\label{2.8}
R_{abcd}(\gamma )=\frac{R(\gamma )}{n(n-1)}\,(\gamma _{ac}\,\gamma _{bd}-
\gamma _{ad}\,\gamma _{bc})\ .
\end{equation}
Hence the slices $z={\rm constant}$ have constant curvature $R(\gamma )/
n(n-1)$ provided $n>2$. The Einstein tensor calculated with $\gamma _{ab}$ 
is given by
\begin{equation}\label{2.9}
G_{ab}(\gamma )=\left (\frac{2-n}{2n}\right )\,R(\gamma )\,\gamma _{ab}\ .
\end{equation}
The Einstein tensor calculated with the metric $g_{\alpha\beta}$ of $M$ has 
non--identically vanishing components
\begin{eqnarray}\label{2.10}
G^a_b&=&{\rm e}^{-2V}\,G^a_b(\gamma )+(n-1)\,\left [V''+\frac{n}{2}\,V'^2\right ]\,\delta ^a_b\ ,\\
G^n_n&=&-\frac{1}{2}\,{\rm e}^{-2V}\,R(\gamma )+\frac{1}{2}\,n(n-1)\,V'^2\ .
\end{eqnarray}
Substituting (\ref{2.9}) into (\ref{2.10}) and eliminating $R(\gamma )$ from the 
resulting equation using (2.11) we arrive at
\begin{equation}\label{2.12}
G^a_b=\left [\frac{1}{2}\,(n-2)\,G^n_n+(n-1)\,(V''+V'^2)\right ]\,\delta ^a_b\ .
\end{equation}
With these preliminaries completed we shall now consider the two cases of the 
history of a domain wall in a vacuum space--time and in (A)dS space--time.

For a domain wall in a vacuum $(G^\alpha _\beta =0)$ the above 
equations give the following equation for $V(z)$:
\begin{equation}\label{2.13}
V''+V'^2=0\ .
\end{equation}
The general solution is 
\begin {equation}\label{2.14}
V(z)=\ln (c\,z+1)\ ,
\end{equation}
where $c$ is a constant of integration. The second constant of integration has 
been put equal to unity without loss of generality. However a planar domain 
wall with $z=0$ as its history in the space--time $M$ must be reflection 
symmetric in $z=0$. To achieve this we adjust (\ref{2.14}) to read
\begin{equation}\label{2.15}
V(z)=\ln (c\,|z|+1)\ ,
\end{equation}
which now, of course, no longer satisfies (\ref{2.13}) on $z=0$ but 
satisfies
\begin{equation}\label{2.16}
V''+V'^2=2\,c\,\delta (z)\ ,
\end{equation}
where $\delta (z)$ is the Dirac delta function which is singular on $z=0$. In 
addition the Einstein tensor of $M$ no longer vanishes throughout $M$. The 
non--identically vanishing components are
\begin{equation}\label{2.17}
G^a_b=2\,c\,(n-1)\,\delta (z)\,\delta ^a_b\ .
\end{equation} 
It thus follows that $z=0$ is the history in $M$ of a domain wall with surface 
energy density $\sigma$ and surface pressure $P$ given by
\begin{equation}\label{2.18}
\sigma =-P=-\frac{c}{4\,\pi}\,(n-1)\ .
\end{equation}
In order to have $\sigma >0$ we must require $c<0$. It follows now from (2.10) 
that
\begin{equation}\label{2.19}
G^a_b(\gamma )=-\lambda\,\delta ^a_b\ ,
\end{equation}
with
\begin{equation}\label{2.20}
\lambda =\frac{1}{2}\,(n-1)(n-2)\,c^2>0\ .
\end{equation}
Hence each hypersurface $z={\rm constant}$ in $M$ is isometric to 
de Sitter space.

Let us now put $x^a=(x^0, x^i)$ with $x^0=t$ and $i=1, 2, 3,\dots ,n-1$. The 
line--element of $M$ takes the form
\begin{equation}\label{2.21}
ds^2=(c\,|z|+1)^2\left [-dt^2+{\rm e}^{-2ct}\,dx^i\,dx^i\right ]+dz^2\ .
\end{equation}
If we transform $z$ to a new transverse coordinate $Z$ on either side of $z=0$ 
such that $c\,|z|+1={\rm e}^{c\,|Z|}$ then this line--element becomes 
\begin{equation}\label{2.22}
ds^2={\rm e}^{2c\,|Z|}(-dt^2+dZ^2)+{\rm e}^{-2c\,(t-|Z|)}\,dx^i\,dx^i\ .
\end{equation}
This generalises to $(n+1)$--dimensional space--time the solution of 
Vilenkin \cite{Vil}. The non--identically vanishing components of the 
Einstein tensor of $M$ now read
\begin{equation}\label{2.23}
G^a_b=2c\,(n-1)\,\delta (Z)\,\delta ^a_b\ .
\end{equation}
Now make the Lorentz boost in the $Z$--direction,
\begin{equation}\label{2.24}
t=\bar t\,\cosh\chi -\bar Z\,\sinh\chi\ ,\qquad Z=\bar Z\,\cosh\chi 
-\bar t\,\sinh\chi\ ,\qquad \bar x^i=x^i\ .
\end{equation}
The Lorentz factor of the boost is $\cosh\chi$. It is convenient to 
rewrite this transformation in terms of the two null coordinates 
$\bar u=\bar t-\bar Z$ and $\bar v=\bar t+\bar Z$ so that 
\begin{equation}\label{2.25}
t=\frac{\bar u}{2}\,{\rm e}^\chi +\frac{\bar v}{2}\,{\rm e}^{-\chi}\ ,
\qquad Z=-\frac{\bar u}{2}\,{\rm e}^{\chi}+\frac{\bar v}{2}\,{\rm e}^{
-\chi }\ .
\end{equation}
Following \cite{AS} this boost is first applied to the line--element 
(\ref{2.22}) and then we take the limit in which the velocity of the 
boost tends to the speed of light ($\chi \rightarrow +\infty $). This 
results in (\ref{2.22}) becoming
\begin{equation}\label{2.26}
ds^2=-{\rm e}^{2\bar c\,|\bar u|}\,d\bar u\,d\bar v+{\rm e}^{-2\bar c\,(
\bar u-|\bar u|)}\,d\bar x^i\,d\bar x^i\ ,\end{equation}
where the parameter $\bar c$ is related to the parameter $c$ in (2.22) 
via
\begin{equation}\label{2.27}
\bar c=c\,\cosh\chi\ .
\end{equation}
The transformed Einstein tensor $\bar G_{\alpha\beta}$ vanishes except 
for 
\begin{equation}\label{2.28}
\bar G_{\bar u\bar u}=2\,\bar c\,(n-1)\,\delta (\bar u)\ .
\end{equation}
When $\bar u\neq 0$ the line--element (\ref{2.26}) can be transformed 
into Minkowskian form
\begin{equation}\label{2.29}
ds^2=-dU\,dV+dX^i\,dX^i\ ,
\end{equation}
by the transformation
\begin{eqnarray}\label{2.30}
U&=&\frac{\epsilon}{2\,\bar c}\{{\rm e}^{2\,\epsilon\,\bar c\bar u}-1\}\ ,\\
V&=&\bar v-\bar c\,(1-\epsilon )\,{\rm e}^{(1-\epsilon )\,\bar c\bar u}
\,X^i\,X^i\ ,\\
\bar x^i&=&X^i\,{\rm e}^{-(1-\epsilon )\,\bar c\bar u}\ ,
\end{eqnarray}
where $\epsilon ={\rm sign}\,(\bar u)$. We conclude from (\ref{2.28}) that 
an Aichelburg--Sexl boost of a planar domain wall in the direction transverse 
to the wall results in a planar null shell whose history is the null 
hypersurface $\bar u=0$ in otherwise flat space--time and with energy 
density $\bar\sigma =\bar c\,(n-1)/4\pi$. The null shell is not accompanied 
by an impulsive gravitational wave since we started with a space--time $M$ 
which was conformally flat.

We next consider a domain wall in (A)dS space--time. Thus in addition to 
our assumption that the space--time $M$ be conformally flat we now assume 
that its Einstein tensor has the form
\begin{equation}\label{2.32}
G^{\alpha}_{\beta}=-\Lambda\,\delta ^{\alpha}_{\beta}\ ,
\end{equation}
where $\Lambda$ is a constant. Using (\ref{2.12}) we find that the equation 
for $V(z)$ in this case reads
\begin{equation}\label{2.33}
V''+V'^2=-\frac{2}{n(n-1)}\,\Lambda\ .
\end{equation}
We obtain reflection symmetric solutions, in the same way as in the 
vacuum case, by first solving this equation and then replacing $z$ by 
$|z|$ in the solution. We find that if $\Lambda >0$ then 
\begin{eqnarray}\label{2.34}
{\rm e}^V&=&\alpha\,\cos A_{+}|z|+\beta\,\sin A_{+}|z|\ ,\\
A_{+}&=&\left [\frac{2\,\Lambda}{n(n-1)}\right ]^{1/2}\ ,
\end{eqnarray}
and if $\Lambda <0$ then
\begin{eqnarray}\label{2.36}
{\rm e}^V&=&\alpha\,\cosh A_{-}|z|+\beta\,\sinh A_{-}|z|\ ,\\
A_{-}&=&\left [-\frac{2\,\Lambda}{n(n-1)}\right ]^{1/2}\ ,
\end{eqnarray}
with $\alpha$ and $\beta$ real constants. In place of (\ref{2.33}) these quantities satisfy
\begin{equation}\label{2.37}
V''+V'^2=-\frac{2}{n(n-1)}\,\Lambda +2\,A_{\pm}\frac{\beta}{\alpha}\,\delta (z)\ ,
\end{equation}
and 
\begin{equation}\label{2.38}
G^a_b=-\Lambda\,\delta ^a_b+2\,A_{\pm}(n-1)\,\frac{\beta}{\alpha}\,\delta (z)\,
\delta ^a_b\ .\end{equation}
In order to have a positive surface energy density we 
must take $\alpha\,\beta <0$. In addition we find in this case that
\begin{equation}\label{2.39}
G^a_b(\gamma )=-\lambda\,\delta ^a_b\ ,
\end{equation}
with
\begin{equation}\label{2.40}
\lambda =\frac{1}{2}\,(n-1)(n-2)A^2_{\pm}(\beta ^2\pm\alpha ^2)\ ,
\end{equation}
and so the hypersurfaces $z={\rm constant}$ have de Sitter geometry provided 
$|\beta |>|\alpha |$. As in the vacuum case we next make a change of the 
coordinate $z$ to $Z$ such that
\begin{equation}\label{2.41}
dZ={\rm e}^{-V}\,dz\ .
\end{equation}
Calling ${\cal V}(Z)=V(z(Z))$ we find that
\begin{equation}\label{2.42}
{\rm e}^{-{\cal V}(Z)}=
\frac{\sqrt{\beta ^2\pm\alpha ^2}\,\cosh\left (A_{\pm}
\sqrt{\beta ^2\pm\alpha ^2}\,
|Z|\right )-\beta\,\sinh\left (A_{\pm}\sqrt{\beta ^2\pm\alpha ^2}\,|Z|\right )}
{\alpha\,\sqrt{\beta ^2\pm\alpha ^2}}\ 
,\end{equation}
and the line--element of $M$ now reads
\begin{equation}\label{2.43}
ds^2=-{\rm e}^{2\,{\cal V}(Z)}(-dt^2+dZ^2)+
{\rm e}^{2\,{\cal V}(Z)+2\,A_{\pm}\sqrt{\beta ^2\pm\alpha ^2}\,t}\,
dx^i\,dx^i\ ,
\end{equation}
where, as in the vacuum case, $x^0=t$. Making the Lorentz boost as before 
and taking the limit $\chi \rightarrow +\infty$ we obtain
\begin{equation}\label{2.44}
ds^2=-{\rm e}^{2\,\bar {\cal V}(\bar u)}\,d\bar u\,d\bar v+
{\rm e}^{2\,\bar {\cal V}(\bar u)-2\,\bar A_{\pm}\sqrt{\beta ^2\pm\alpha ^2}\,
\bar u}\,
d\bar x^i\,d\bar x^i\ ,
\end{equation}
with 
\begin{equation}\label{2.45}
\bar A_{\pm}=A_{\pm}\,\cosh\chi\ ,\qquad \bar\Lambda =\Lambda\,\cosh ^2\chi\ ,
\end{equation}
and 
\begin{equation}\label{2.46}
{\rm e}^{-\bar {\cal V}(\bar u)}=
\frac{\sqrt{\beta ^2\pm\alpha ^2}\,\cosh\left (\bar A_{\pm}
\sqrt{\beta ^2\pm\alpha ^2}\,
|\bar u|\right )-\beta\,\sinh\left (\bar A_{\pm}
\sqrt{\beta ^2\pm\alpha ^2}\,|\bar u|\right )}
{\alpha\,\sqrt{\beta ^2\pm\alpha ^2}}\ 
.\end{equation}
The Einstein tensor in the infinite boost limit vanishes except for 
\begin{equation}\label{2.47}
\bar G_{\bar u\bar u}=-2\,\bar A_{\pm}(n-1)\,\frac{\alpha\,\beta}{|\alpha |}\,
\delta (\bar u)\ .
\end{equation}
Here again we have arrived at a null shell in flat space--time.

\setcounter{equation}{0}
\section{Boosting Cosmic Strings}\indent 

A cosmic string is a line source characterized by a conical singularity.
For an infinite straight cosmic string at rest and extending along the 
$z$-axis, the metric takes the form
\begin{equation}\label{3.1}
ds^2=-dt^2 + d\rho^2 + a^2 \rho^2 d\phi^2 + dz^2 \, .
\end{equation} 
Here, $\phi$ runs from $0$ to $2\pi$, and the parameter $a$ takes
account of the conical deficit $\Delta \phi$, given by
\begin{equation}\label{3.2}
\frac{\Delta \phi}{2\pi} = 1-a = {4 \mu} \, ,
\end{equation}
where $\mu$ is the mass per unit length of the string. In terms of the rectangular co-ordinates
\begin{equation}\label{3.3}
x =\rho \cos \phi \qquad , \qquad y = \rho \sin \phi \, ,
\end{equation}
the metric (\ref{3.1}) reads
\begin{equation}\label{3.4} 
ds^2=-dt^2 + dx^2 + dy^2 + dz^2 - (1-a^2)\frac{(xdy-ydx)^2}{x^2+y^2} \, .
\end{equation}

We now boost the string sideways along the $x$-direction, 
by making the co-ordinate transformation
\begin{eqnarray}\label{3.5}
t=\bar t\,\cosh\chi -\bar x\,\sinh\chi=\frac{\bar u}{2}\,{\rm e}^\chi +\frac{\bar v}{2}\,{\rm e}^{-\chi}\, ,\nonumber\\
x=\bar x\,\cosh\chi -\bar t\,\sinh\chi=-\frac{\bar u}{2}\,{\rm e}^{\chi}+\frac{\bar v}{2}\,{\rm e}^{-\chi }\, ,\nonumber\\
y=\bar y\ \qquad ,\qquad z=\bar z\ \, .
\end{eqnarray}
Here $\bar u=\bar t-\bar x$, $\bar v=\bar t+\bar x$ are the plane 
light--like co-ordinates.
The flat part of the line element (\ref{3.4}) (the first four terms) then
becomes
\begin{equation}\label{3.6}
-d\bar ud\bar v+ d\bar y^2 +d\bar z^2 \, .
\end{equation}
In the limit ($\chi \rightarrow +\infty$), the boost (\ref{3.5}) becomes
light--like. On noting that 
\begin{eqnarray}\label{3.7}
x dy - y dx =  
\frac{e^{\chi}}{2}(\bar y d\bar u -\bar u d\bar y) + O(e^{-\chi})\, ,
\nonumber\\
x^2 + y^2 = \frac{e^{2\chi}}{4}(\bar u^2 + b^2)  + O(e^{-2\chi})\ ,
\end{eqnarray}
where 
\begin{equation}\label{3.8}
b^2  = 4\, e^{-2\chi}(\bar y^2 - \frac{1}{2}\bar u \bar v)\, ,
\end{equation}
the last term of (\ref{3.4}) reduces to
\begin{equation}\label{3.9}
-(1-a^2)\frac{(\bar y d\bar u - \bar u d\bar y)^2}{\bar u^2 + b^2}  
+ O(e^{-2\chi})\, .
\end{equation}
The linear mass density of the string measured in the laboratory frame is
\begin{equation}\label{3.10}
\bar \mu = \mu \cosh \chi = 
\frac{1}{4} (1-a) \cosh \chi \, .
\end{equation}
To obtain a regular limiting geometry we must require $\bar \mu$ 
to remain bounded as $\chi \rightarrow +\infty$. This introduces an additional
factor $e^{-\chi}$ into (\ref{3.9}) through the coefficient $(1-a^2)$. 
From the identity
\begin{equation}\label{3.11}
\lim_{b\longrightarrow 0} \frac{b}{\bar u^2 +b^2}=\pi \delta (\bar u)\, ,
\end{equation}
it follows by (\ref{3.8}) that
\begin{equation}\label{3.12}
\lim_{\chi \longrightarrow +\infty}
\frac{e^{-\chi}}{\bar u^2 +b^2}=\frac{\pi}{2|\bar y|}\delta(\bar u)\, .
\end{equation}
Inserting this into (\ref{3.9}) and combining with (\ref{3.6}), we obtain
the final limiting form of the boosted metric
\begin{equation}\label{3.13}
d\bar s^2= \lim_{\chi \longrightarrow +\infty}ds^2=
-d\bar u d\bar v + d\bar y^2 + d\bar z^2 -8\pi \bar \mu |\bar y|
\delta(\bar u)\, d\bar u^2\, .
\end{equation}
This represents the geometry of an infinite straight light--like string 
whose world-sheet is the $2$-flat
\begin{equation}\label{3.14}
\bar u = \bar t  -\bar x =0\, \qquad , \qquad \bar y =0\, ,
\end{equation}
and whose mass per unit length measured in the laboratory frame is $\bar \mu$.
 
We shall now show that (\ref{3.13}) describes a conical singularity
along the $\bar z$-axis moving at the speed of light in the $\bar x$-
direction, with angular deficit $\bar \Delta \phi$ related to the line
density $\bar \mu$ in the laboratory frame by an equation identical in form 
to (\ref{3.2}).
There are two ways of approach to this. We can start from the
conical curvature singularity of metric (\ref{3.1}) representing
the original time--like string in its rest-frame and apply the
boost (\ref{3.5}). Alternatively we can work directly from (\ref{3.13}). 
This is a metric of Kerr-Schild form. Both approaches are of interest and we shall consider them in turn.

In  the first approach we begin with the static string line-element
(\ref{3.1}) or (\ref{3.4}). The angular deficit $\Delta \phi$
of a $2$-plane ${\cal S}_2$ of constant $z$ and $t$ is associated with
the distributional curvature
\begin{equation}\label{3.15}
^{(2)}R=2 (\Delta \phi)\,  \delta_2\, ,
\end{equation}
where
\begin{equation}\label{3.16}
\delta_2 =\, ^{(2)}g^{-1/2}\,\delta (x)\, \delta (y) = 
\frac{1}{a}\,\delta (x)\, \delta (y)\ ,
\end{equation}
is the invariant two-dimensional delta-function concentrated at the origin.
To check (\ref{3.15}), note that for Gaussian polar co-ordinates
\begin{equation}\label{3.17}
ds^2_{(2)} = d\rho ^2 + f^2(\rho,\phi)d\phi ^2\ ,
\end{equation}
and the Gaussian curvature is
\begin{equation}\label{3.18}
^{(2)}R = -\frac{2}{f}\, \partial^2_{\rho}f(\rho,\phi) \, .
\end{equation}
We temporarily smooth out the conical singularity at $\rho = 0$ by
choosing $f$ to be any smooth function $f_{\epsilon}(\rho)$
satisfying the conditions
\begin{equation}\label{3.19}
f_{\epsilon}(\rho)=a\rho \qquad (\rho \geq \epsilon)\qquad , \qquad
\lim_{\rho \longrightarrow 0}\frac{f_{\epsilon}(\rho)}{\rho}=1\, .
\end{equation}
Then
\begin{equation}\label{3.20}
\int \int \,\,^{(2)}R\, ^{(2)}g^{1/2} d\rho d\phi = 
-4\pi \int^{\epsilon}_0 (\partial ^2_{\rho} f_{\epsilon}) d\rho =
4\pi (1-f'_{\epsilon}(\epsilon)) = 4\pi (1-a)\, .
\end{equation}
The conical deficit $\Delta \phi$ is defined by
\begin{equation}\label{3.21}
2\pi - \Delta \phi = 
\lim_{\epsilon \longrightarrow 0}\left (\frac{{\rm circumference}}
{{\rm radius}}\right )_{\rho=\epsilon}
=\lim_{\epsilon \longrightarrow 0}\frac{f_\epsilon (\epsilon)}{\epsilon}
=2\pi a\, .
\end{equation}
Combining these results leads to (\ref{3.15}).

The Ricci tensor of the $xy$-plane ${\cal S}_2$ is, by (\ref{3.15}),
\begin{equation}\label{3.22}
 ^{(2)}R_{ab} = (\Delta \phi )\, ^{(2)}g_{ab}\, \delta_2\, ,
\end{equation}
in which the $2$-metric $^{(2)}g_{ab}$ can be decomposed as
\begin{equation}\label{3.23}
^{(2)}g_{ab}=\rho ,_a\rho ,_b + a^2 \rho^2 \phi ,_a\phi ,_b\, .
\end{equation}
Since the extrinsic curvature of  ${\cal S}_2$ is zero, the
four-dimensional Ricci tensor of the metric (\ref{3.1}) can be read off
at once from (\ref{3.22}). It is
\begin{equation}\label{3.24}
^{(4)}R_{\alpha \beta}=\frac{\Delta \phi}{a}\, \delta (x)\, \delta (y)\,
(\rho,_\alpha \rho,_\beta + a^2 \rho^2  \phi,_\alpha \phi,_\beta)\, , 
\end{equation}
in which Greek indices are $4$-dimensional and a comma denotes 
partial differentiation.

Under the boost (\ref{3.5}), we have for large $\chi$,
\begin{equation}\label{3.25}
\rho \simeq |x| \simeq \frac{e^\chi}{2}|\bar u|\ , \qquad
\phi=\tan^{-1}\frac{\bar y}{\bar x}=O(e^{-\chi})\ ,\qquad
\delta (x) \simeq 2 e^{-\chi} \delta (\bar u) \ .
\end{equation}
We rescale $\Delta \phi$ to a new parameter $\bar \Delta \phi$ by
analogy with (\ref{3.10}),
\begin{equation}\label{3.26}
\bar \Delta \phi = (\Delta \phi ) \cosh \chi \, ,
\end{equation}
and require $\bar \Delta \phi$ to stay bounded as $\chi \rightarrow +\infty$.
In this limit the second term of (\ref{3.24}) becomes negligible, 
$a \rightarrow 1$ by (\ref{3.2}), and we obtain
\begin{equation}\label{3.27}
^{(4)}R_{\alpha \beta}= ( \bar \Delta \phi)\, \delta (\bar u)\, 
\delta (\bar y)\,
\bar u,_\alpha \,\bar u,_\beta \, .
\end{equation}
Thus the stress-energy tensor
\begin{equation}\label{3.28}
T_{\alpha\beta}=\frac{ \bar \Delta \phi}{8\pi}\, \delta (\bar u)\, 
\delta (\bar y)\, \bar u,_\alpha\, \bar u,_\beta \, ,
\end{equation}
is indeed that of a distributional light--like source having as its history 
in space--time the null
$2$-flat $\bar u =\bar y =0$, and with mass per unit length
\begin{equation}\label{3.29}
\bar \mu = \frac{\bar \Delta \phi}{8\pi}\ , 
\end{equation}
measured in the laboratory frame.

To establish the role of $ \bar \Delta \phi$ in (\ref{3.27}) as an
angular deficit in the laboratory frame, consider the $2$-space $\bar {\cal S}_2$
of constant $\bar z$ and $\bar t$ 
in the geometry (\ref{3.13}). 
Recalling that $\bar u = \bar t - \bar x$, we obtain from (\ref{3.27}),
\begin{equation}\label{3.30}
^{(4)}R_{\bar x \bar x}= (\bar \Delta \phi)\, 
\delta (\bar x)\, \delta (\bar y)\, .
\end{equation}
Now, $\bar u,_\alpha$ is light--like and geodesic for the flat background
of the Kerr-Schild metric (\ref{3.13}), and therefore retains these 
properties with respect to the full metric. From this geodesic
property and the Gauss-Codazzi equations it follows that 
${}^{(4)}R_{\bar x \bar x}={}^{(2)}R_{\bar x \bar x}$. Thus
\begin{equation}\label{3.31}
\frac{1}{2}\, ^{(2)}R =\,^{(2)}R_{\bar x \bar x}=\,^{(4)}R_{\bar x \bar x}=
(\bar \Delta \phi)\, \delta (\bar x)\,  \delta (\bar y) \, .
\end{equation}
According to (\ref{3.15}), this is indeed the distributional curvature in the
$\bar x \bar y$-plane corresponding to an angular deficit 
$\bar \Delta \phi$.

The alternative route to these results is to proceed directly from the
Kerr-Schild metric (\ref{3.13}) for the light--like string. Starting with 
the line--element 
\begin{equation}\label{3.32}
d\bar s^2 = -d\bar u d\bar v + d\bar y^2 + d\bar z^2 + 
2 H(\bar u,\bar v,\bar y,\bar z)\, d\bar u^2\ ,
\end{equation}
the Ricci tensor is
\begin{equation}\label{3.33}
^{(4)}R_{\alpha \beta}= (8HH,_{\bar v \bar v} -\nabla^2 H)\bar u,_\alpha \,
\bar v_\beta -4H,_{\bar v \bar v} \, \bar u,_{(\alpha}\bar v,_{\beta )}
-4H,_{\bar v a} e^{(a)}_{(\alpha}\, \bar u,_{\beta )}\, ,
\end{equation}
where $\nabla^2 = \partial^2_{\bar y} + \partial^2_{\bar z}$, 
$\bar x^a =(\bar y, \bar z)$ and $e^{(a)}_\alpha = \partial \bar x^a\,/\,
\partial \bar x^\alpha$. Inserting $H=-4\pi \tilde \mu \, \delta (\bar y)\, \delta (\bar u )$
from (\ref{3.13}), so that 
$\nabla^2 H = - 8\pi \bar \mu \, \delta (\bar y)\, \delta (\bar u )$,
we obtain
\begin{equation}\label{3.34}
^{(4)}R_{\alpha \beta}= 8\pi \bar \mu\, \delta (\bar y)\, \delta (\bar u )\,
\bar u,_\alpha\, \bar u,_\beta\, ,
\end{equation}
in agreement with (\ref{3.27}) and  (\ref{3.29}).

\section*{Acknowledgment}\noindent
This collaboration has been funded in part by the Minist\`ere des Affaires 
\'Etrang\`eres, D.C.R.I. 220/SUR/R and by a NATO Collaborative Linkage
Grant (CLG.976417).

\end{document}